\documentstyle[epsfig,epsf]{article}
\def\s2t {\sin^2 2\theta}

\begin{document}

\begin{center}
{\Large {\bf Bologna and the Cooperation with Developing Countries}}
\end{center}

\vskip .5 cm

\begin{center}
Giorgio Giacomelli \\
{\it University of Bologna and INFN, Sezione di Bologna, \\
V.le C. Berti Pichat 6/2, I-40127 Bologna, Italy\\} 

E-mail: giacomelli@bo.infn.it 

\par~\par

Presented at the Meeting "Scientific Cooperation between Italy and
Developing Countries: The Program TRIL-ICTP",    
\\ ENEA, Lungotevere Thaon de Revel 76, Roma, 10 March 2006. 

\vskip .7 cm
{\large \bf Abstract}\par
\end{center}

{\normalsize A brief recollection is made of the many scientific
collaborations of Bologna with colleagues from Developing Countries. In
particular are discussed in detail the collaborations in some recent 
AstroparticlePhysics Experiments.}

\vspace{5mm}
 
\large

\section{Introduction}\label{sec:intro}
	Immediately after World War 2 the list of Developing Countries
included also Japan and several European Nations. During this period,
relatively small experiments were made in fundamental High Energy
Physics, mainly in the US and later at CERN. In many of these
experiments there were young physicists from ``Developing Countries",
mainly on an individual basis [1-5]. But soon after larger
Collaborations started and some were officially formalized, like the
CERN-Serpukhov Collaboration [3], several bubble chamber collaborations
and experiments at the new CERN facilities [6-9].\par
It was during this period of time that Italy started to have
medium-large facilities and that ICTP started scientific programs, like
the TRIL program (Training and Research in Italian Laboratories). ICTP had
always organized Summer Schools, but during this period their number
increased considerably and some of the subjects of the School became
more technical and more experimental. And it was always a pleasure to
have lunch in the ICTP cafeteria with Abdus Salam, the Director and
founder of ICTP, and Giuseppe Furlan, the leader of the TRIL program.\par
Later the Italian National Institutes (INFN, ENEA, CNR, ...) started to
provide funds FAI, for international visitors. And the University of
Bologna, like other Italian Universities, started to have
Internationalization Programs which involved also many Developing
Nations. Several agreements were signed between Universities and also
between Institutes/Departments.\par
In Bologna we started having long visits of colleagues from Bolivia,
Brazil, India Mali, Morocco, Nepal, Pakistan, Romania and Syria, and
shorter visits of colleagues from Algeria, Colombia, India, Japan,
Morocco, Russia, Spain and several other Nations. Our group had in Bologna 
at any
moment a couple of visitors for long stages and a couple of visitors for
relatively short visits. And some research groups from Morocco, Romania,
Mali started to participate officially in large non accelerator
astroparticle physics experiments, mainly at the Gran Sasso National
Laboratories [10-14] and also in other places [15-18]. 

\section{Scientific fields}
	Early scientific experimental collaborations concerned
experiments at small and large accelerators, mainly in the field of
hadron-hadron collisions [1, 5, 7]. They were performed to study the low
energy parameters of pion-nucleon scattering, to search for resonances
and threshold phenomena in relatively low energy $\pi^{\pm}N$,
$k^{\pm}N$, $p^{\pm}N$ total cross
sections and in $\pi^{\pm}A$ absorption cross sections. At larger energies a
topic of great interest was the discovery of the rising of the total 
hadron-hadron cross
sections, first $K^+p$ at Serpukhov [3] and those of the other cross sections 
at CERN and at Fermilab [4, 5]; other topics of interest were 
low $p_t$ and high $p_t$ 
particle production. Large bubble chamber experiments allowed to study 
several fields, 
including resonance
searches, studies of exotic channels, and neutrino-nucleon deep
inelastic scattering [7]. \par
The more recent collaborations involved Non Accelerator Astroparticle
Physics experiments, mainly performed at the Gran Sasso National
Laboratories, but also at high altitude laboratories. Some people and 
groups from Developing Nations participated actively in the MACRO experiment 
at the Gran Sasso underground laboratory [10-14], in the SLIM experiment 
at the Chacaltaya high altitude laboratory [16] and in the CAKE 
balloon experiment [15]. \par
In the largest experiments there was also involvement in Detector
Development and Construction [9, 12, 17, 18]. Detector development and
use of these detectors to study also environmental parameters was made,
in particular in the collaborations with ENEA [19].

\section{Theses, Summer Schools, Workshops}
The agreements between the University of Bologna and
Universities in Developing Countries allowed many PhD theses to be
performed, based on the results of common experiments, like MACRO at the
Gran Sasso Lab, SLIM at the Chacaltaya High Altitude Laboratory in Bolivia
and in the CAKE balloon experiment for the study of the primary Cosmic
Ray Composition. In the last 15 years a total of about 2 Laurea theses
and 10 PhD theses were made.\par
	The involvement of Moroccan Colleagues in the MACRO experiment
allowed to perform a NATO-Advanced Research Workshop in Oujda, Morocco
(``Cosmic Radations: from Astronomy to Particle Physics") [20]. This
allowed to get together about 50 physicists and astrophysicists involved
in non-accelerator astroparticle physics experiments in general and in
the field of neutrino oscillations in particular. The participants were
from 11 Nations. This encouraged new relations and new cultural
exchanges. Also many contacts were established with Institutions from
Mediterranean Countries. Also a first small experiment was made to
measure for the first time the radon contamination in the Oujda region
(see ref. [20] page 325).\par
	A series of 7 Schools on ``Non Accelerator Astroparticle Physics"
were held at the Trieste Abdus Salam International Center for
Theoretical Physics; the proceedings of the last four Schools were
published, usually by World Scientific Publishing, see for ex. the
proceedings of the 7th School [21]. Fig. 1 shows the beginning of a
short report in the CERN Courier on this 7th School [22]. Notice that
about 80 students from 43 countries attended the last School and about
16 renowned physicists lectured at the School.

\begin{figure}
\begin{center}
{\centering\resizebox*{!}{7.8 cm}{\includegraphics{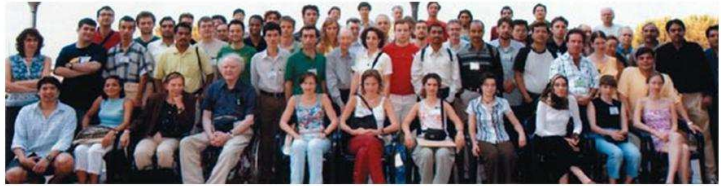}}\par}
\begin{quote}
\caption {The $7^{th}$ School on ``Non-Accelerator Astroparticle Physics", as 
reported in the CERN Courier (22)} 
%\label{fig:soudan2_detector}
\end{quote}
\end{center}
\end{figure}

\section{Conclusions. Outlook}
	The ICTP programs have been very useful for young people from
Developing Countries and encouraged them to join the experiments performed
in Italy, in CERN and in other laboratories. Other ICTP programs also helped the 
young people, after their
return in their home Countries, to keep in touch with their European
collaborators. Other programs, the numerous summer schools, the books
donation program, the transport contributions, etc. have been also very
important.\par 
The financial contribution from the FAI funds and from fellowships for
foreigners by INFN, ENEA, CNR and ASI, were also successfully used.

\section{Acknowledgements}
	I would like to thank ICTP for their many programs which involve
people from Developing Countries and for organizing all sorts of Summer
Schools. I thank ENEA, CNR and INFN for providing funds FAI and
fellowships for foreigners, and the University of Bologna for making
many collaborative agreements with Universities in Developing Countries.
I thank A. Casoni for typing this manuscript.\\

{\normalsize In the early collaborations I like to recall the colleagues Kozo 
Miyake from Japan, B. A. Leontic from Yugoslavia, J. Derkaoui from Morocco, R.
Sosnowsky and others from Poland, S. P. Denisov and others from the
USSR, J. Negret and others from Colombia, Hi Wang and others from China.
For the more recent: collaborations organized via ICTP, INFN, ENEA and the
University of Bologna, I recall the colleagues Eduardo Medinaceli from
Bolivia, Eudice Vilela from Brazil, Bipasha Pal, Jayta Gosh and Ashavani
Kumar from India, Vincent Togo and Abdramane Ba from Mali, Jamal Derkaoui,
Hassane Dekhissi, Dikra Bakari, Fatiha Maaroufi, Mohamed Ouchrif, 
Abdeslem Rrhioua, Abdelilah Moussa
and Taoufic Ouali from Morocco, Zouleikha Sahnoun from Algeria, B. B. Bam
from Nepal, Shahid Manzoor and Ikram Shahzad from Pakistan, Vlad Popa,
Marius Rujoiu, George Serbanut and Dimitru Hasegan from Bucarest, 
Valeri Tioukov from Russia.

}
\end{document}